\documentclass[]{spie}  %>>> use for US letter paper
\usepackage{amsmath,amsfonts,amssymb}
\usepackage{graphicx}
\usepackage[colorlinks=true, allcolors=blue]{hyperref}
	\usepackage[perpage]{footmisc} %restart footnote number on each page
	\usepackage{amsmath}
	\usepackage{mathrsfs} %mathy script letters
	\usepackage{amssymb}
	\usepackage{amsfonts}
	\usepackage{graphicx}  
	\usepackage{setspace}
	\usepackage{cancel}
    \usepackage[left=0.875in,right=0.875in,top=1in,bottom=1.25in]{geometry}
	\usepackage{enumitem}
	\usepackage{etoolbox}
	\usepackage{siunitx}	
	\usepackage{multirow,bigdelim}
	\usepackage{textcomp}
	\usepackage{float}
	\usepackage{color, colortbl} %need for color tables like from Excel2Latex
	\usepackage{booktabs} %for making fancy lines in tables like Excel2Latex
	\usepackage{caption}
	\usepackage[font=singlespacing]{caption}
	\usepackage{wasysym}
	\usepackage{pdflscape}
	\usepackage{afterpage}
	\usepackage{url}
    \usepackage{upgreek}
	\usepackage{lscape} %lets you make landscape pages with \begin{landscape}
	\usepackage{subcaption} %lets you make subfigures
	\usepackage{mathtools}          %loads amsmath as well

	\AtBeginEnvironment{quote}{\singlespacing\small\it}
	\usepackage{xspace}
    \newcommand{\um}{$\upmu$m\xspace}	%xspace makes it not leave a weird space if there's punctuation after the command	
	\renewcommand{\deg}{$^{\circ}$ }

	\newcommand{\planck}{\emph{Planck }}
    \newcommand{\arcsec}{$^{\prime\prime}$}	
    \newcommand{\arcmin}{$^{\prime}$}
	\setlist[enumerate]{label*=\arabic*.}
	\setlist[enumerate,1]{label*=\arabic*.,font=\bfseries,before=\bfseries}
	\setlist[enumerate,2]{label*=\arabic*.,font=\normalfont,before=							\normalfont}
	\setlistdepth{9}

\title{Design and characterization of a balloon-borne diffraction-limited submillimeter telescope platform for BLAST-TNG}

\author[a]{Nathan P. Lourie}
\author[a]{Francisco E. Angil\'e}
\author[b]{Peter C. Ashton}
\author[c]{Brian Catanzaro}
\author[a]{Mark J. Devlin}
\author[a]{Simon Dicker}
\author[d]{Joy Didier}
\author[e]{Bradley Dober}
\author[f]{Laura M. Fissel}
\author[g]{Nicholas Galitzki}
\author[h]{Samuel Gordon}
\author[a]{Jeffrey Klein}
\author[a]{Ian Lowe }
\author[h]{Philip Mauskopf}
\author[a]{Federico Nati}
\author[i]{Giles Novak}
\author[j]{L. Javier Romualdez}
\author[k]{Juan D. Soler}
\author[i]{Paul A. Williams}

\affil[a]{University of Pennsylvania, 209 South 33rd St, Philadelphia, USA}
\affil[b]{University of California - Berkeley, Lawrence Berkeley National Laboratory, 1 Cyclotron Rd, CA, USA}
\affil[c]{CFE Services, 5147 Pacifica Dr, San Diego, USA}
\affil[d]{University of Southern California, Los Angeles,
USA}
\affil[e]{National Institute of Standards and Technology, 325 Broadway, Boulder, USA}
\affil[f]{National Radio Astronomy Observatory, 520 Edgemont Rd,
Charlottesville, U.S.A}
\affil[g]{University of California San Diego, 9500 Gilman Dr, La Jolla, USA}
\affil[h]{School of Earth and Space Exploration, Arizona State University, Tempe, USA}
\affil[i]{Center  for  Interdisciplinary  Exploration  and  Research  in Astrophysics  and  Department  of  Physics  \&  Astronomy,  Northwestern University,  2145 Sheridan  Road, USA}
\affil[j]{University of Toronto Institute for Aerospace Studies, 4925 Dufferin Street, Toronto, Canada}
\affil[k]{Max-Planck-Institute for Astronomy, Konigstuhl 17, Heidelberg, Germany.}

\authorinfo{Further author information: (Send correspondence to N.P.L)\\N.P.L.: E-mail: nlourie@sas.upenn.edu, Telephone: 1 215 573 5445}

\begin{document} 
\maketitle

\begin{abstract}
The Next Generation Balloon-borne Large Aperture Submillimeter Telescope (BLAST-TNG) is a submillimeter mapping experiment planned for a 28 day long-duration balloon (LDB) flight from McMurdo Station, Antarctica during the 2018-2019 season.  BLAST-TNG will detect submillimeter polarized interstellar dust emission, tracing magnetic fields in galactic molecular clouds. BLAST-TNG will be the first polarimeter with the sensitivity and resolution to probe the $\sim$0.1 parsec-scale features that are critical to understanding the origin of structures in the interstellar medium. With three detector arrays operating at 250, 350, and 500 \um (1200, 857, and 600 GHz), BLAST-TNG will obtain diffraction-limited resolution at each waveband of 30, 41, and 59 arcseconds respectively.

To achieve the submillimeter resolution necessary for its science goals, the BLAST-TNG telescope features a 2.5 m aperture carbon fiber composite primary mirror, one of the largest mirrors flown on a balloon platform. Successful performance of such a large telescope on a balloon-borne platform requires stiff, lightweight optical components and mounting structures. Through a combination of optical metrology and finite element modeling of thermal and mechanical stresses on both the telescope optics and mounting structures, we expect diffraction-limited resolution at all our wavebands. We expect pointing errors due to deformation of the telescope mount to be negligible. We have developed a detailed thermal model of the sun shielding, gondola, and optical components to optimize our observing strategy and increase the stability of the telescope over the flight. We present preflight characterization of the telescope and its platform.

\end{abstract}


\section{Introdution}
\label{section:intro}

The Next Generation Balloon-borne Large Aperture Submillimeter Telescope (BLAST-TNG) is a submillimeter mapping experiment which features three microwave kinetic inductance detector (MKID) arrays operating over 30\% bandwidths centered at 250, 350, and 500 \um (1200, 857, and 600 GHz). These highly-multiplexed, high-sensitivity arrays, featuring 918, 469, and 272 dual-polarization pixels, for a total of 3,318 detectors, are coupled to a 2.5 m diameter primary mirror and a cryogenic optical system providing diffraction-limited resolution of 30\arcsec, 41\arcsec, and 50\arcsec \space respectively. The arrays are cooled to $\sim$275 mK in a liquid-helium-cooled cryogenic receiver which will enable observations over the course of a 28-day stratospheric balloon flight from McMurdo Station in Antarctica as part of NASA's long-duration-balloon (LDB) program, planned for the 2018/2019 winter campaign. BLAST-TNG is the successor to the BLASTPol and BLAST balloon-borne experiments which flew five times between 2005 and 2012\cite{enzo_blast,fissel_blastpol}.

Achieving diffraction-limited, sub-arcminute resolution and telescope pointing accuracy is one of the highest priorities for the success of the BLAST-TNG mission. Although the science goals of BLAST-TNG are similar to the 2012 BLASTPol mission, most of the major instrument systems have been rebuilt and improved since the last flight. A new 2.5 m aperture Cassegrain telescope, featuring a lightweight composite carbon fiber reinforced polymer (CFRP) primary mirror designed and built by Alliance Spacesystems,\footnote{4398 Corporate Center Dr, Los Alamitos, CA 90720} will enable an increase in resolution to 30\arcsec \space at 250 \um, from BLASTPol's 2.5\arcmin \space at the same band. With improved detector sensitivity and a increase in detector count by a factor of 12, we expect BLAST-TNG will have more than six times the mapping speed of BLASTPol. The new cryostat has demonstrated a 28 day hold-time, enabling observations of many more targets at greater depth than were possible during the $\sim$13 day BLASTPol flight in 2012.

The primary science goal of BLAST-TNG is to map the polarized thermal emission from galactic interstellar dust around star-forming regions and in the diffuse interstellar medium (ISM). These maps will yield $\sim$250,000 polarization vectors on the sky, allowing us to explore correlations between the magnetic field dispersion, polarization fraction, cloud temperature, and column density. Quantifying the relationships between these variables over a large sample of clouds will yield testable relationships which can be fed back into numerical simulations. The \planck satellite has observed strong correlations between the orientation of Galactic magnetic fields and large-scale ISM structures \cite{planck_XXXII}, as well as the interior of giant molecular clouds (GMCs)\cite{planck_XXX}. While BLASTPol was able to observe the magnetic fields within GMCs at higher resolution than \planck \cite{fissel_blastpol,soler_blastpol}, BLAST-TNG will be the first experiment to probe the fields within the characteristic filamentary structures within GMCs observed by \emph{Herschel} \cite{hill}. Combining the BLAST-TNG data with molecular cloud simulations, \cite{soler_sims_2013}  and numerical models of dust emission \cite{guillet_2018} and  grain properties, \cite{draine_hensley} will give unprecedented insight into the interplay between the gravitational, turbulent and magnetic field contributions to star and cloud formation, as well as the physics of grain alignment and mass flow within the interstellar medium.

Polarized dust emission is also the dominant foreground for observations of the cosmic microwave background (CMB). Characterization of these foregrounds is one of the most important requirements in the search for the gravitational wave signature of cosmic inflation \cite{cmb_s4}. While the power spectrum from polarized dust foregrounds is thought to be lowest at small angular scales, there is limited high-resolution observational data of the diffuse ISM \cite{planck_XXX,caldwell_EEBB}.  BLAST-TNG will be able to make the deepest maps to date of the dust emission in the types of dark, diffuse regions of the sky favored by state of the art CMB polarization experiments. BLAST-TNG will probe angular scales not well-characterized to date, and explore correlations between diffuse dust emission and structures in the cold neutral medium \cite{ghosh} at submillimeter wavelengths where the intensity of the thermal dust signal dominates.  With its high pixel count and photon-noise-limited detectors, BLAST-TNG will produce maps of diffuse ISM with higher fidelity than the highest frequency \emph{Planck} polarization maps at 353 GHz.

\section{BLAST-TNG Composite Optics Technology}
In order to meet our angular resolution requirements, BLAST-TNG must feature large-aperture, lightweight telescope mirrors. Building-large aperture balloon-borne telescope optics is particularly challenging. With inadequate support, gravitationally-induced sag can introduce serious aberrations, yet support structures must be small enough to fit on NASA launch vehicles, and light enough not to compromise altitude during flight. Achieving diffraction-limited performance in the submillimeter requires highly accurate optics with rms wavefront errors of order $\sim$10\um. Traditional mirror fabrication techniques are inadequate to meet all the requirements for BLAST-TNG within the cost restrictions of a balloon mission. To date, most balloon payloads operating in the millimeter/submillimeter wavebands have used aluminum mirrors, but have been limited by mechanical constraints to less than 2 m in diameter, including BLASTPol (1.8 m) \cite{enzo_blast} and EBEX (1.5 m) \cite{ebex}. The SOFIA instrument features the largest sub-orbital primary mirror to date, at 2.7 m \cite{sofia}, made out of Zerodur, a ceramic silicate material that is lighter and stiffer than aluminum. However, while SOFIA's 808 kg primary mirror is well-suited for a large airplane, it is unacceptably heavy for a balloon experiment. Low-density metals, such as the beryllium alloy used in the construction of the JWST mirrors \cite{jwst} are extremely expensive. Mirrors made from these types of materials are cost-effective only if they are able to be reused for multiple flights, making them risky to use on a balloon mission where they may not be recovered at all. 

The CFRP mirror which will fly on BLAST-TNG represents a significant technological development and research effort. CRFP composites have many advantages over traditional metal mirrors. They have a strength-to-weight ratio many times that of aluminum, and a near-zero coefficient of thermal expansion, which is especially desirable for a balloon platform, as the thermal environment in flight can be unstable. Composite primary mirrors are produced via replication from a positive mold. High-surface accuracy molds are much cheaper to produce than lightweight mirrors, and can be reused to make subsequent mirrors with equivalent optical quality without additional polishing. While composite materials are expensive, the reduced recurring costs for replicating extremely lightweight mirrors make them well-suited to balloon experiments which plan on making repeated flights with no guarantee of recovering the telescope components.

Large-aperture composite mirrors with high surface accuracy have been demonstrated in the submillimeter. The first launch of the BLAST experiment featured a 2 m composite primary mirror developed for the \emph{Herschel} space telescope \cite{catanzaro_blast05}, although its performance in flight was degraded due to a lack of active focusing control.  With a 2.5 m aperture, the BLAST-TNG primary mirror will be both the largest mirror ever flown on a balloon experiment, and the largest CFRP telescope mirror operating at submillimeter wavelengths (THz frequencies). This mirror was designed in partnership with a commercial collaborator, Alliance Spacesystems, under a NASA Small Business Innovation Research (SBIR) grant. 
\section{BLAST-TNG Optical Architecture}
\label{section:overview}

\begin{figure}[!htb]
\includegraphics[width=\linewidth]{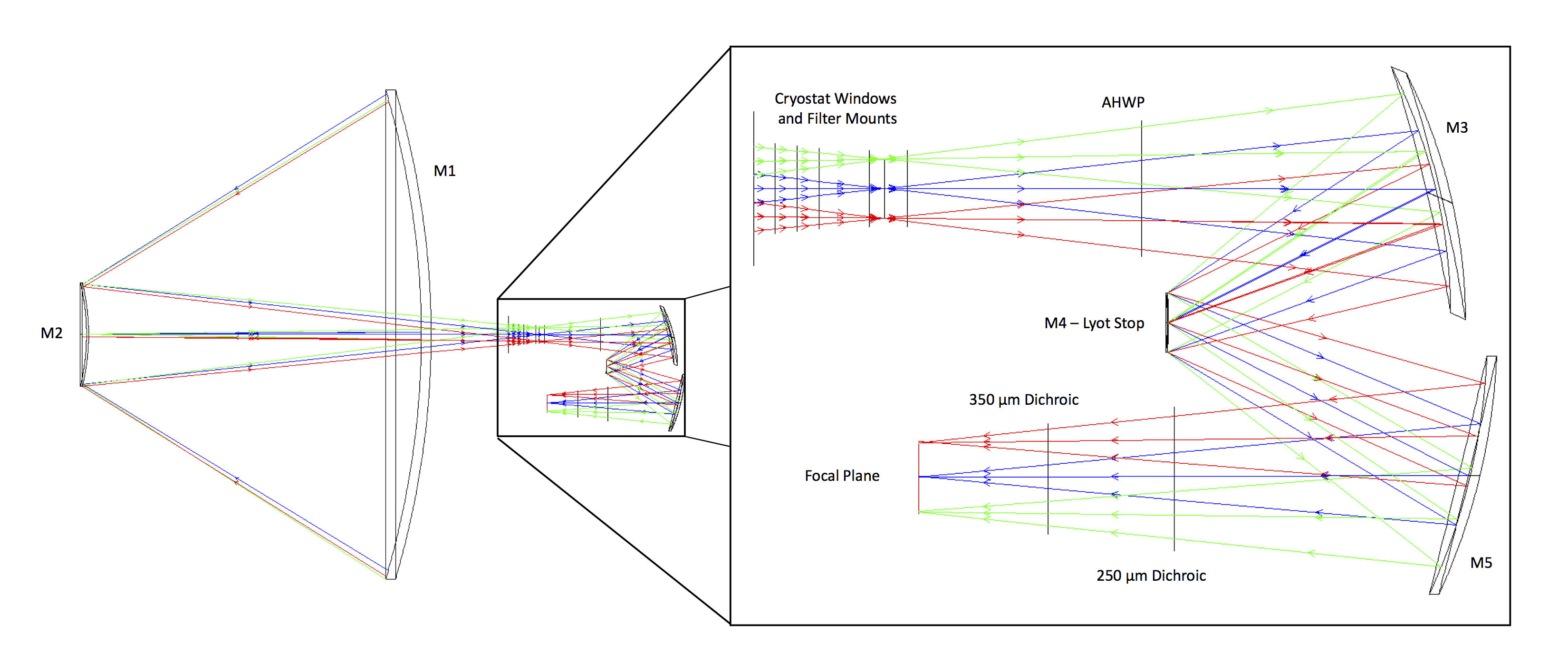}
\caption[BLAST-TNG Optics Design]{Ray trace of the BLAST-TNG optics design from Zemax design software. The left-hand side shows the on-axis Cassegrain telescope formed by the primary (M1) and secondary mirrors (M2). The Cassegrain focus lies within the 4 K optics box, shown in the small rectangle, as well as the blown up inset. Light enters the optics box towards the top left side of the enclosure where it passes through the window of the cryogenic receiver and a series of filters. After these filters, the first optical element is the broadband achromatic half wave plate (AHWP), followed by the modified Offner relay formed by the three mirrors M3, M4 (the Lyot stop), and M5. The location of the three focal plane arrays are shown schematically.} 
\label{figure:optics_design}
\end{figure}

The BLAST-TNG optics design is based on a 2.5 m aperture on-axis Cassegrain telescope, with a CFRP composite primary mirror and an aluminum secondary mirror. The secondary mirror is mounted on three linear actuators which can move the secondary in piston/tip/tilt to account for changes in telescope focus due to differential thermal contraction of the telescope mirrors, CFRP support struts, and the aluminum gondola. The optical design is shown in Fig. \ref{figure:optics_design}.

The telescope feeds a cold (4 K) reimaging optics system which refocuses the beam onto three focal plane arrays. The cold optics are arranged in a modified Offner relay configuration, shown in the inset of Fig. \ref{figure:optics_design}. A similar configuration was flown in the BLAST/BLASTPol optics box. This configuration has several advantages, namely (1) it is compact, a necessary condition for running the optics in a liquid-helium-cooled cryostat, (2) the main optical elements all lie in a single plane, allowing all the elements to be mounted to a single sturdy optics bench, (3) the modified relay can be used to illuminate the focal planes with a different F/\# than the telescope beam, and (4) the cryogenic Lyot stop allows us to limit the illumination of the primary mirror which reduces the thermal loading on the detectors. The cold optics simultaneously illuminate three focal plane arrays of Microwave Kinetic Inductance Detectors, which are optically coupled via single-mode feedhorns. The feedhorns were designed and machined at Arizona State University, based on a modified Potter design \cite{potter}, and were drilled from a monolithic aluminum block with custom-manufactured drill bits. Details of the cryogenic receiver design can be found in refs. \citenum{lourie_cryo,tyr_blasttng_spie}.

\begin{figure}[!htb]
\includegraphics[width=\linewidth]{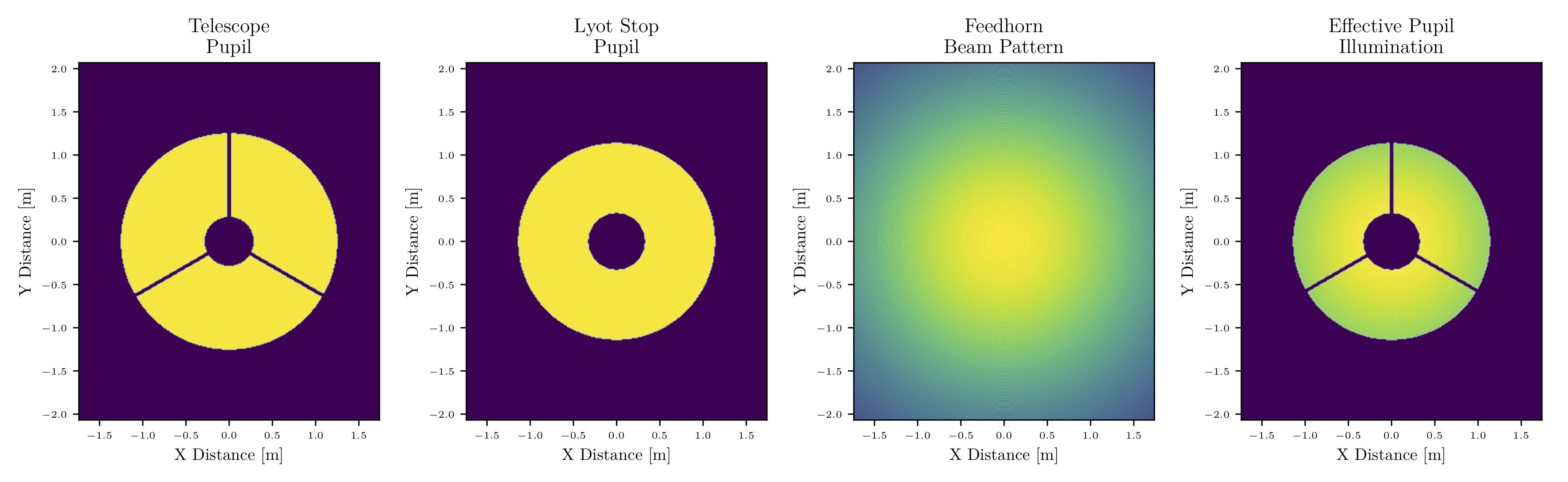}
\caption[Effective Pupil Function] {Diagrams of the telescope pupil including obscuration from the struts and secondary mirror, the image of the Lyot stop on the primary mirror, the near-Gaussian illumination of the primary mirror by the Feedhorns. The illumination of the effective system pupil is shown at right.}
\label{figure:pupil}
\end{figure}

The cold Lyot stop is placed at the image of the primary mirror and acts as the limiting aperture of the system. The Lyot stop limits the central beam illumination of the primary mirror to 2.33 m. The detector feedhorns provide a near-Gaussian beam which overfills the Lyot stop, and tapers the illumination by 4.6 dB at the edge of the Lyot to reduce ringing in the beam. The illumination of the system pupil is shown in Fig. \ref{figure:pupil}. While all of the feedhorn beams from each of the focal plane arrays overlap on the Lyot stop, they do not illuminate the same area of the primary mirror. A summary of the BLAST-TNG telescope optical design is given in Table \ref{table:optical_prescription}. 

% Table generated by Excel2LaTeX from sheet 'Telescope Optical Prescription'
\begin{table}[htbp]
  \renewcommand{\arraystretch}{1.5} % space the rows a bit more
  \centering
  \caption{BLAST-TNG Telescope Optical Prescription}
    \begin{tabular}{crl}
    \toprule

    \textbf{Element} & \multicolumn{1}{c}{\textbf{Parameter}} & \multicolumn{1}{c}{\textbf{Value}} \\
        
    \midrule
    \midrule
    \multirow{3}[2]{*}{Primary Mirror} & Clear Aperture & \multicolumn{1}{r}{2500.0 mm} \\
          & Radius of Curvature & \multicolumn{1}{r}{4132.10766 mm} \\
          & Conic Constant & \multicolumn{1}{r}{-1.000} \\
    \midrule
    \multirow{3}[2]{*}{Secondary Mirror} & Clear Aperture & \multicolumn{1}{r}{573.0 mm} \\
          & Radius of Curvature & \multicolumn{1}{r}{1209.82622 mm} \\
          & Conic Constant & \multicolumn{1}{r}{-2.380136} \\
    \midrule
    \multirow{7}[2]{*}{Telescope} & \multicolumn{1}{p{8.25em}}{Primary Vertex to Secondary Vertex} & \multicolumn{1}{r}{1590.0 mm} \\
          & EFL   & \multicolumn{1}{r}{9698.82 mm} \\
          & F/\#  & \multicolumn{1}{r}{3.87953} \\
          & FOV   & \multicolumn{1}{r}{23 arcminutes} \\
          & Obscuration Ratio & \multicolumn{1}{r}{7.871\%} \\
          & Strut Obscuration Ratio & \multicolumn{1}{r}{2.618\%} \\
    \bottomrule
    \end{tabular}%
  \label{table:optical_prescription}%
\end{table}%

\section{Optical Requirements}
\label{section:optics_reqs}

\subsubsection{Telescope Optical Requirements}

The BLAST-TNG telescope must maintain diffraction-limited performance under all combined stresses and loading conditions throughout the anticipated 28-day LDB flight. The telescope must be operational on the ground for pre-flight integration and characterization, and must operate in flight over a broad range of temperatures and pointing angles. 

To ensure diffraction-limited performance, the driving requirement for the telescope design was that the total wavefront error (WFE) of the telescope be no greater than 10 \um \space rms under all combined loading conditions. This requirement would ensure that the telescope Strehl ratio be no less than $\approx$ 90\% across all wavebands. A summary of the telescope performance specifications is given in Table \ref{table:telescope_specs}.

There are three major loads that the telescope was designed to operate under. These loads were studied extensively with finite element modeling (FEM) trade studies which ultimately drove the final design of the telescope optics and support system:
\begin{enumerate}

\item{Gravity Sag}\\
\textnormal{
Gravity-induced sag is the dominant loading stress for the BLAST-TNG telescope. Ground-based telescopes do not have strict limitations on the mass of telescope support structures, and space-borne missions which which have extremely demanding mass limits are not affected by gravity sag at all. Sub-orbital flights occupy the worst of both worlds -- having both gravitational stresses and mass constraints. 
}

\item{Thermal Soak/Gradients}
\textnormal{
During the flight, the telescope will operate with a thermal shroud or Sun shield which should help control the thermal environment and block direct solar illumination. Even so, we anticipate large thermal gradients across the structure, in particular between the primary and secondary mirrors. These gradients were measured to be up to 15-20 $^\circ$C during the 2010 and 2012 BLASTPol flights. To account for this, the ability to refocus the telescope in flight is critical. 
}

\item{Hygroscopic Strain}
\textnormal{
The resin in CFRP composites exhibits temperature-dependent absorption of water vapor. The amount of water vapor absorbed by the composite depends on the relative humidity (RH) of the environment. The rate at which the moisture content of the composite changes is inversely proportional to the environment temperature \cite{moisture}. As the mirror absorbs (desorbs) water it will grow (shrink) changing the radius of curvature and the conic constant similar to a change in temperature. 
}
\end{enumerate}

\begin{table}[htbp]
  \renewcommand{\arraystretch}{1.5} % space the rows a bit more

  \centering
  \caption{BLAST-TNG Telescope Performance Specifications}
    \begin{tabular}{ll}
	\toprule
    
    \multicolumn{1}{c}{\textbf{Parameter}} & \multicolumn{1}{c}{\textbf{Value}} \\
    \midrule
    \midrule
    Telescope Wavefront Error & $\leq$ 10 \um \space rms \\
    
    \multirow{3}[0]{*}{Primary Mirror Surface Error } \hspace{1in} \ldelim\{{3}{1mm}&  $\leq$ 7 \um \space rms on 50-250 cm scales \\
          &  $\leq$ 4 \um \space rms on 5-50 cm scales \\
          & $\leq$ 2 \um \space rms on 0-5 cm scales \\ 
    Secondary Mirror Suface Error & $\leq$ 1 \um \space RMS \\
    Operational Elevation Range & 0\deg to 60\deg \\
    Change in Pointing Error from 20\deg to 60\deg & \textless 10 arcseconds \\
    Operating Temperature & -20 $\pm$ 15 \deg C \\
    \multicolumn{1}{p{20.25em}}{Temperature Difference Between Primary and Secondary Mirrors} & \textless 20\deg C \\
    Operating Pressure & 3 mbar \\
    Secondary Mirror In-Flight Positioning & $\delta Z_{min}$ = 10 \um \\
    Total Mass & \textless 150 kg \\
    First Natural Frequency & \textless 35 Hz \\
    Strut Obscuration Ratio & \textless 3 \\% \\
    \bottomrule
    \end{tabular}%
  \label{table:telescope_specs}%
\end{table}%

\section{Telescope Design and Fabrication}
\label{telescope_design}

\begin{figure}[ht!]
\includegraphics[width=\linewidth]{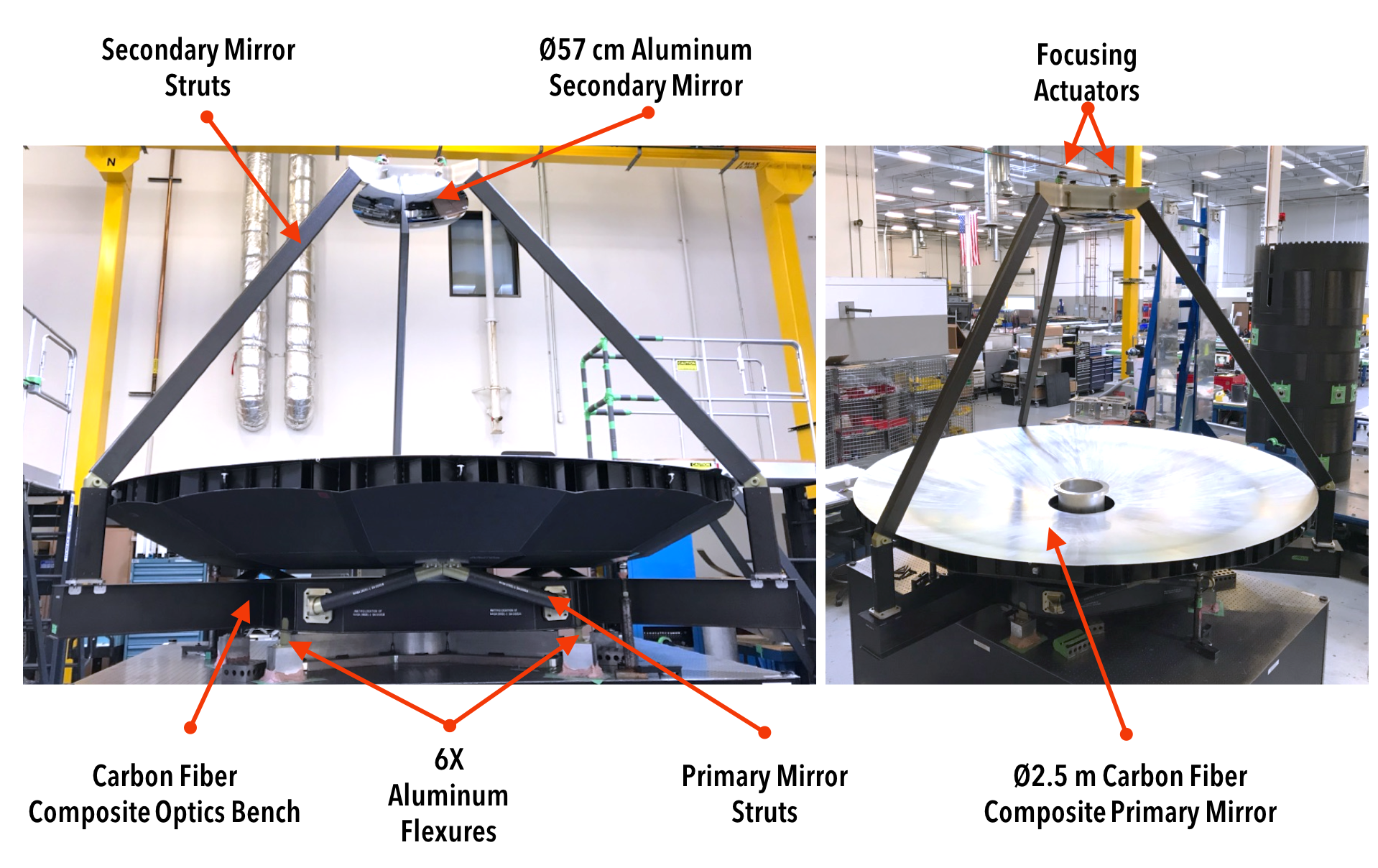}
\caption[BLAST-TNG Telescope Design Overview] {The completed BLAST-TNG telescope during assembly at Alliance Spacesystems, showing the primary and secondary mirrors, the CFRP struts and optical bench, and secondary mirror actuators. The aluminum flexures mount to the aluminum gondola inner frame.}
\label{figure:telescope_redline}
\end{figure}

The BLAST-TNG telescope design is comprised of three major components: the primary and secondary mirrors which form the telescope beam, and a sturdy optical bench. The design of each of these components is detailed in the sections below. The final design of each component is the result of exhaustive finite element modeling, trade studies, and incorporates lessons learned from previous mirror designs. The design team had experience building and designing previous generations of BLAST \cite{enzo_blast} as well as composite optics and reflectors for NASA space missions including \emph{Herschel} \cite{catanzaro_blast_metrology}, WMAP space telescope \cite{wmap_optics}, and MAVEN \cite{maven}. The main elements of the telescope are detailed in Fig. \ref{figure:telescope_redline}.

The optical bench, or reaction structure, supports the primary mirror, the secondary mirror struts, and mounts to the balloon gondola. It is responsible for holding the mirrors in a low-stress, kinematic configuration to maintain precise alignment between the mirrors and the cryogenic receiver while not transmitting bending stresses to the optical surfaces. The optical bench is built of flat composite laminates bonded together in a rigid, box-like structure with internal webbing. The laminates are formed by laying down layers of 0.13 mm-thick, layers of carbon fiber preimpregnated with epoxy, known as prepreg. The laminates are layers of uni-directional tape where all the fibers are oriented in a single direction, as opposed to woven material.  To achieve nearly isotropic stiffness and thermal expansion in the plane of the laminate, the orientation of each layer of the laminate is rotated by 45\deg from the previous layer. The layers are then vacuum-pressed together and heat-cured. The bench itself is roughly 15 cm thick. The strength and rigidity of the bench is proportional to the thickness. In order to increase the thickness of the bench without altering the location of the primary mirror surface, the front facesheet has a cutout so that the rear surface of the primary mirror sits below the front of the bench.

The primary mirror is supported by a pseudo-kinematic mount comprised of three bipods. Each bipod is made of two composite tubes bonded to aluminum fittings on each end. On one end these fittings bolt to aluminum fittings on the side of the optics bench, and on the other are bolted to Invar fittings bonded to the interior webbing of the primary mirror. FEM showed that making the metallic fittings out of Invar was critical for managing the deformation of the primary under thermal stresses. Invar, like CFRP composites, has a low CTE. Bonding higher CTE materials like aluminum directly to the primary mirror caused significant thermal deformation around the bond sites. FEM showed that under the -20 $^\circ$C uniform soak expected during flight, switching from aluminum to Invar greatly reduced the rms surface error from 51.9 to 11.1 \um, and the peak-to-valley surface error from 3.4 to 0.6 \um.

\begin{figure}[ht!]
\centering
\includegraphics[width=.75\linewidth]{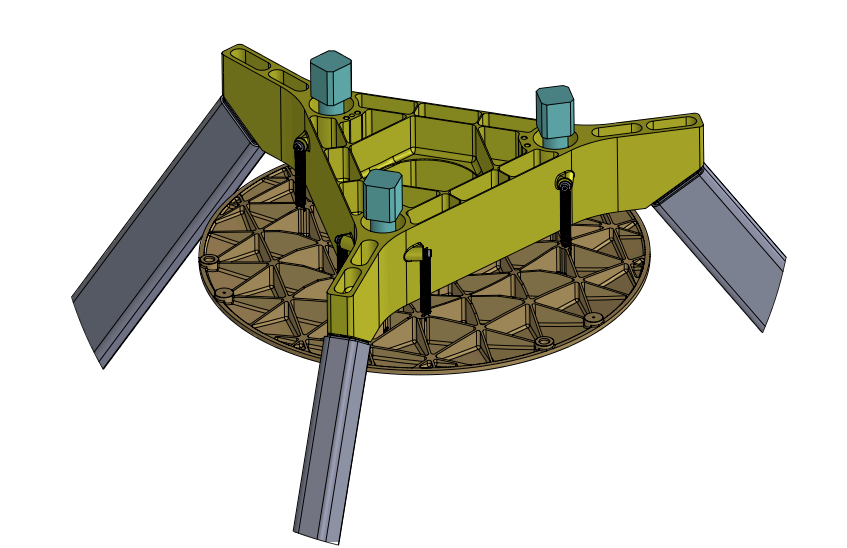}
\caption[Secondary Mirror Assembly Render] {CAD Render of the secondary mirror assembly and support system. The aluminum secondary mirror is shown as a copper color. The push plate (yellow) is the aluminum mirror support which is bolted to the secondary struts (grey). The three focusing actuators are shown in teal. The actuators are mounted to the push plate, and actuate the mirror by pushing against a v-groove block with a ball-end. The mirror is pulled against the ball/groove fittings by the six extension springs shown.}
\label{figure:m2ass}
\end{figure}

The optical bench also supports three long struts which support the secondary mirror assembly. Like the primary mirror struts, the secondary mirror struts are made from all-composite tubes bonded to aluminum end fittings. The strut tubes have a rectangular cross section, with their long sides oriented radially outward from the optical axis. This cross-section provides sufficient stiffness while reducing the obscuration of the primary mirror. The far end of the struts is bolted to an aluminum triangular structure, known as the ``push-plate" which supports the secondary mirror and the focusing actuators. The secondary mirror assembly is shown in Fig. \ref{figure:m2ass}.

\subsection{Primary Mirror}

The BLAST-TNG composite primary mirror features a  monolithic front facesheet which is coated with a thin layer of  vapor-deposited aluminum (VDA). The mirror's structural strength comes from an interior CFRP core. The core is formed by a honeycomb-like composite modules with flat laminate strips forming triangular voids. This isogrid structure provides a high bending modulus while maintaining low mass, both of which minimize the effect of gravitational sag. Additional stiffness is provided by a segmented rear facesheet which helps distribute bending stresses across the core. The material selected for all the structural components was a woven prepreg combining K63712\footnote{Mitsubishi Chemical Carbon Fiber and Composites, Inc.} graphite fiber and BT250E low-temperature-cure (121 $^\circ$C) epoxy prepreg.\footnote{TenCate Advanced Composites} This roughly 60\%/40\% fiber/resin composite provides low CTE and high modulus. 

The modulus of the isogrid core is a strong function of the core thickness. FEM studies led to increasing the core thickness as much as possible given the fixed positions of the primary mirror surface and the gondola mount surface. At its thickest point, roughly half the radius from the optical axis, the core is roughly 30 cm thick. At its thinnest, around the central hole and at outer edge of the mirror, the core is only 6 cm thick. This tapered shape maximizes the strength at the areas of highest bending stress, while reducing mass where the stresses are low.

\begin{figure}[!htb]
\includegraphics[width=\linewidth]{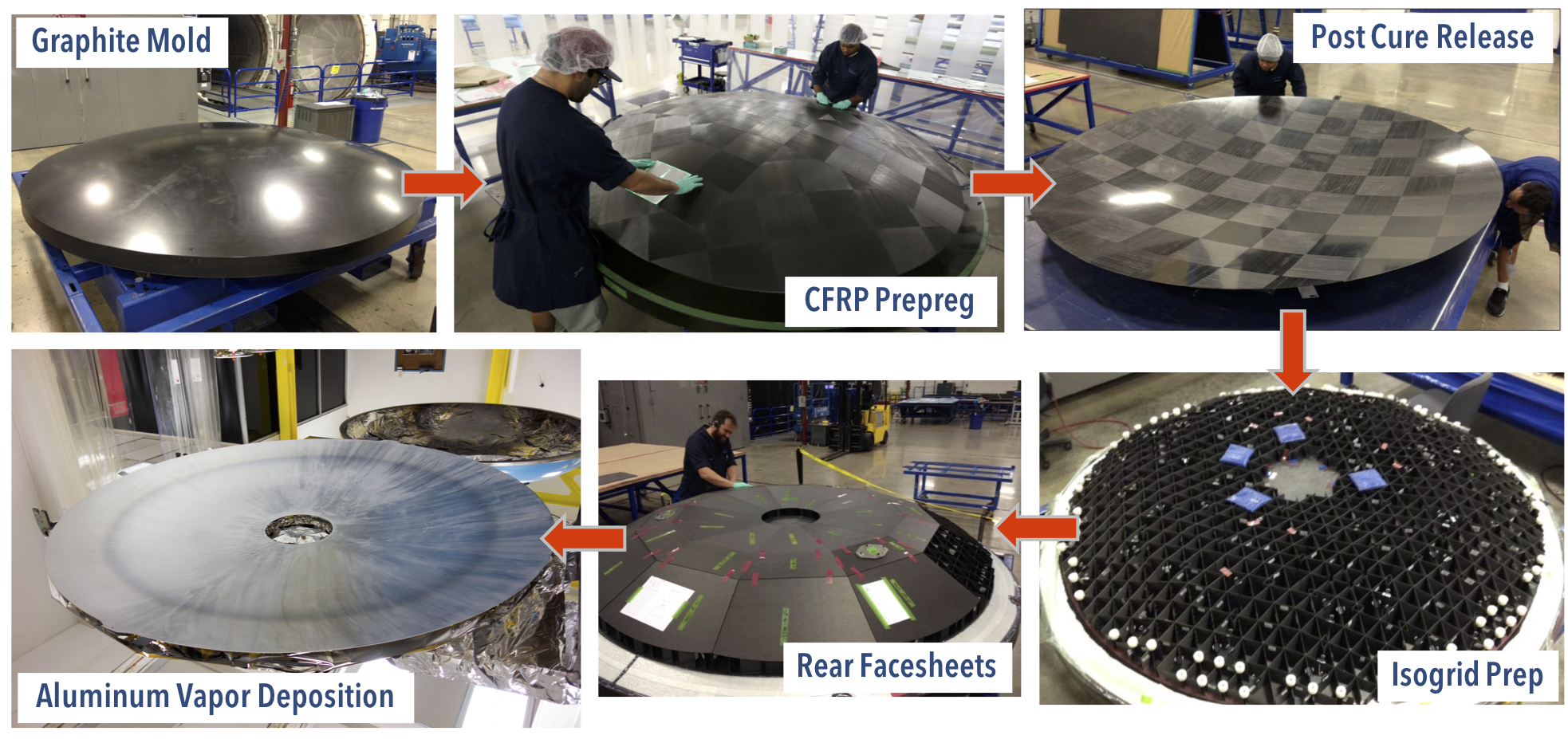}
\caption[Primary Mirror Manufacturing Process] {Primary Mirror Manufacturing Process}
\label{figure:m1_manufacture}
\end{figure}

The optical surface of the primary mirror was made by laying up multiple layers of prepreg onto a positive graphite mold. The mold was rough-machined from graphite using a computer numerically controlled (CNC) mill and then hand polished to the final figure. After the front facesheet was laid up on the mold it was cured in an autoclave while vacuum pressed against the mold. After curing the facesheet was pulled off the mold and inspected. Once inspected the optical surface was placed back on the mold, and the isogrid was built up in small modular sections. The assembly flow for the primary mirror is shown in Fig. \ref{figure:m1_manufacture}.

\subsection{Secondary Mirror \label{subsection:m2_design} }

\begin{figure}[H]
\includegraphics[width=\linewidth]{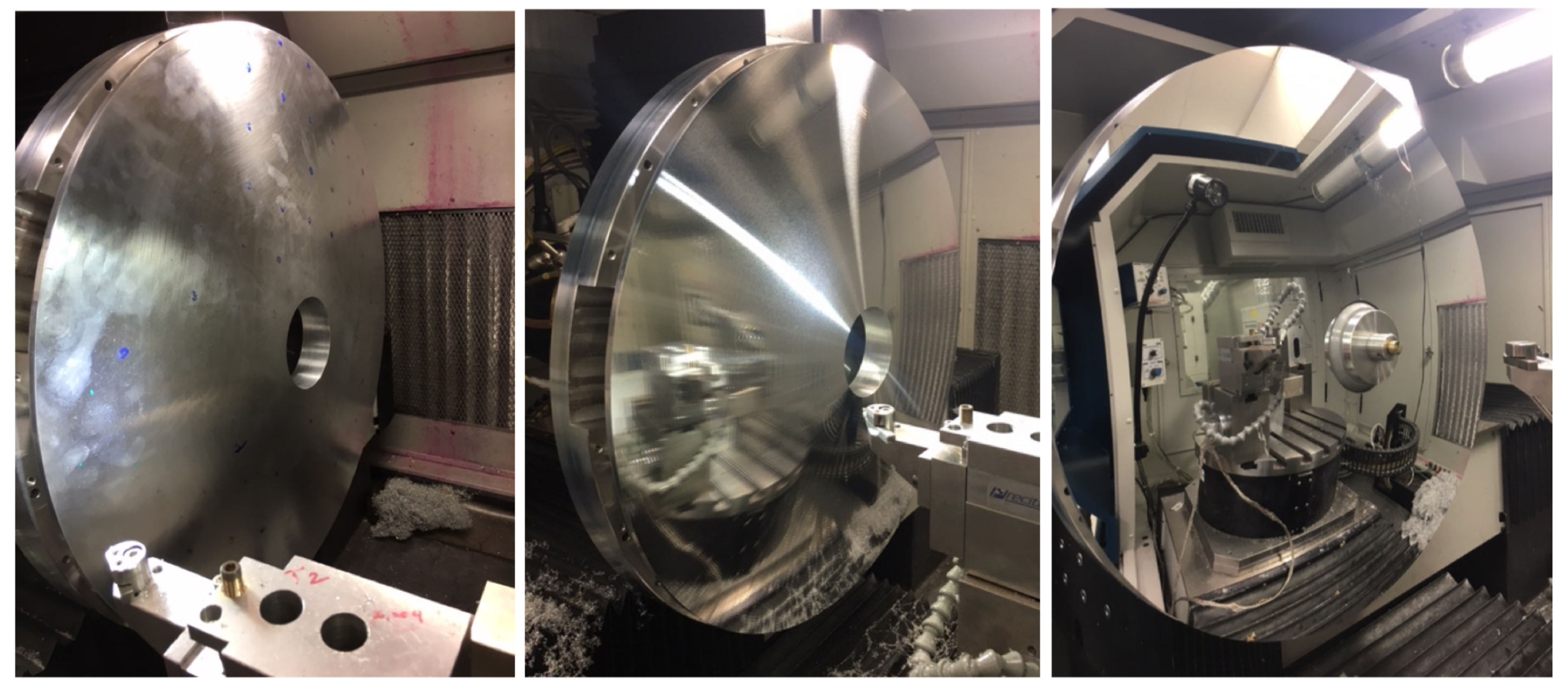}
\caption[Secondary Mirror Diamond Turning Progression] {Images of the secondary mirror at NiPro Optics Inc. taken before (left), during (center), and after (right) diamond turning. After the final machining pass (right) the mirror is nearly optical quality, and produces a clear image of the tool rest.}
\label{figure:m2_turning}
\end{figure}

The comparatively small size of the secondary mirror ($\diameter$ 573 mm) greatly reduced the complexity of the engineering and manufacturing. Lightweight $\sim$0.5 m infrared and submillimeter optics are widely manufactured, and as such we could rely on existing capabilities of different vendors throughout the process. Corning NetOptix Inc.\footnote{69 Island Street, Keene, NH 03431} developed a stiff, aggressively light-weighted design based on preliminary designs by Alliance Spacesystems, and provided two rough-machined mirror blanks.  After surveying the blanks at UPenn, the one with the lower degree of surface figure error was diamond-turned by NiPro Optics Inc.\footnote{7 Marconi, Irvine, CA 92618}.  The progression of the diamond-turning is shown in Figure \ref{figure:m2_turning}.

\section{Expected Performance}
\subsection{Metrology}
Complete characterization of the full BLAST-TNG telescope optical system is only possible during the flight. In-band absorption from water vapor prevents ground-based observations of far-field sources with the flight receiver. To evaluate the focus for the BLAST-Pol experiment, the telescope was refocused to image near-field sources up to a few hundred yards away. While this approach is useful for testing focus routines, the WFE introduced by moving the focus, and the atmospheric absorption make meaningful evaluation of the telescope beam impractical. To characterize the expected telescope performance for BLAST-TNG, we rely mainly on analytical performance models and finite element analysis. Extensive FEM under all loading conditions was performed at Alliance Spacesystems, with optical analysis performed throughout the design process to ensure that the 10 \um rms WFE condition was met under all gravitational, thermal, and hygroscopic stresses expected during operation.

The accuracy of the final primary mirror surface figure presented the largest uncertainty throughout the design process. Because the surface accuracy and size scale of the primary mirror are unprecedented for a composite mirror, there is little heritage for predicting the surface figure errors of the finished surface, and large-scale deviations from the mold surface. Additionally, limited metrology data was available during the intermediate stages of the manufacture and for the final product. Traditional evaluation techniques for large-aperture infrared and submillimeter telescope optics such as infrared interferometry \cite{catanzaro_blast_metrology}, wavefront sensing \cite{shack_hartmann_jwst}, and coordinate measuring machine (CMM) profilometry were all cost-prohibitive given the size of the primary mirror. 

\begin{figure}[!htb]
\includegraphics[width=\linewidth]{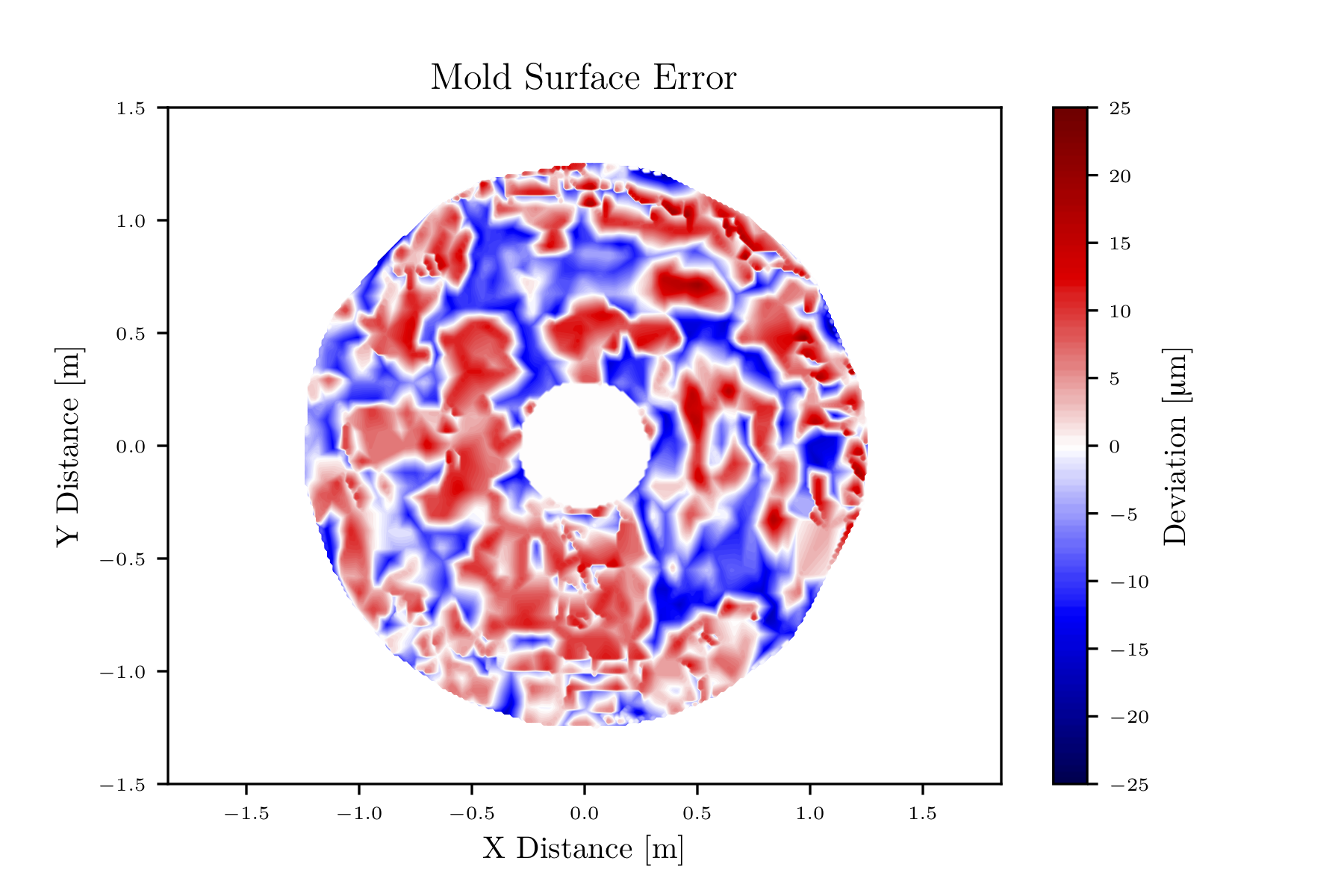}
\caption[M1 Mold Surface Error] {Surface error of the primary mirror graphite mold after final polishing measured using a commercial laser tracker. Deviations shown are calculated after subtracting the best-fit conic section and tip/tilt.}
\label{figure:mold_surface}
\end{figure}%

Both the mold and the primary mirror surface were surveyed using a laser tracker\footnote[0]{FARO Technologies Inc.}, a technique which has been demonstrated at the \textless1 \um accuracy level \cite{laser_tracker}, and surface data recorded on a $\sim$5 cm square grid. The final survey of the mold is shown in Fig. \ref{figure:mold_surface}. After removing the best-fit conic section and tip/tilt of the surface, the mold demonstrates a surface error of 8.9 \um rms, and 40.6 \um peak-to-valley. The optical surface of the primary mirror was also surveyed after initial release from the mold, and twice after the mirror assembly was completed before application of the VDA optical surface. While surface measurements were sufficient for placing an upper limit on the surface figure error, large-spatial-scale surface errors were not repeatably observed between subsequent measurements. We attribute these systematic errors to the relative thermal and mechanical stability of the mirror itself during these measurements, as compared with the mold.

\subsection{Point-Spread Function}
The available metrology data were incorporated into a model of the system optical response based on the nominal optics design and the measured feedhorn beam pattern. The model uses the Zemax model for the full optical system as the nominal reference prescription, models the feedhorn response as a true Gaussian, and assumes that the primary mirror surface perfectly replicates the mold surface. Although the surface of the mold exhibits a number of high frequency surface errors, the impact to performance is minor due to the wavelength compared to the size of the deformations. We also assume that the secondary mirror is refocused to account for deviations from the nominal primary mirror conic section. 

\begin{figure}[H]
\includegraphics[width=\linewidth]{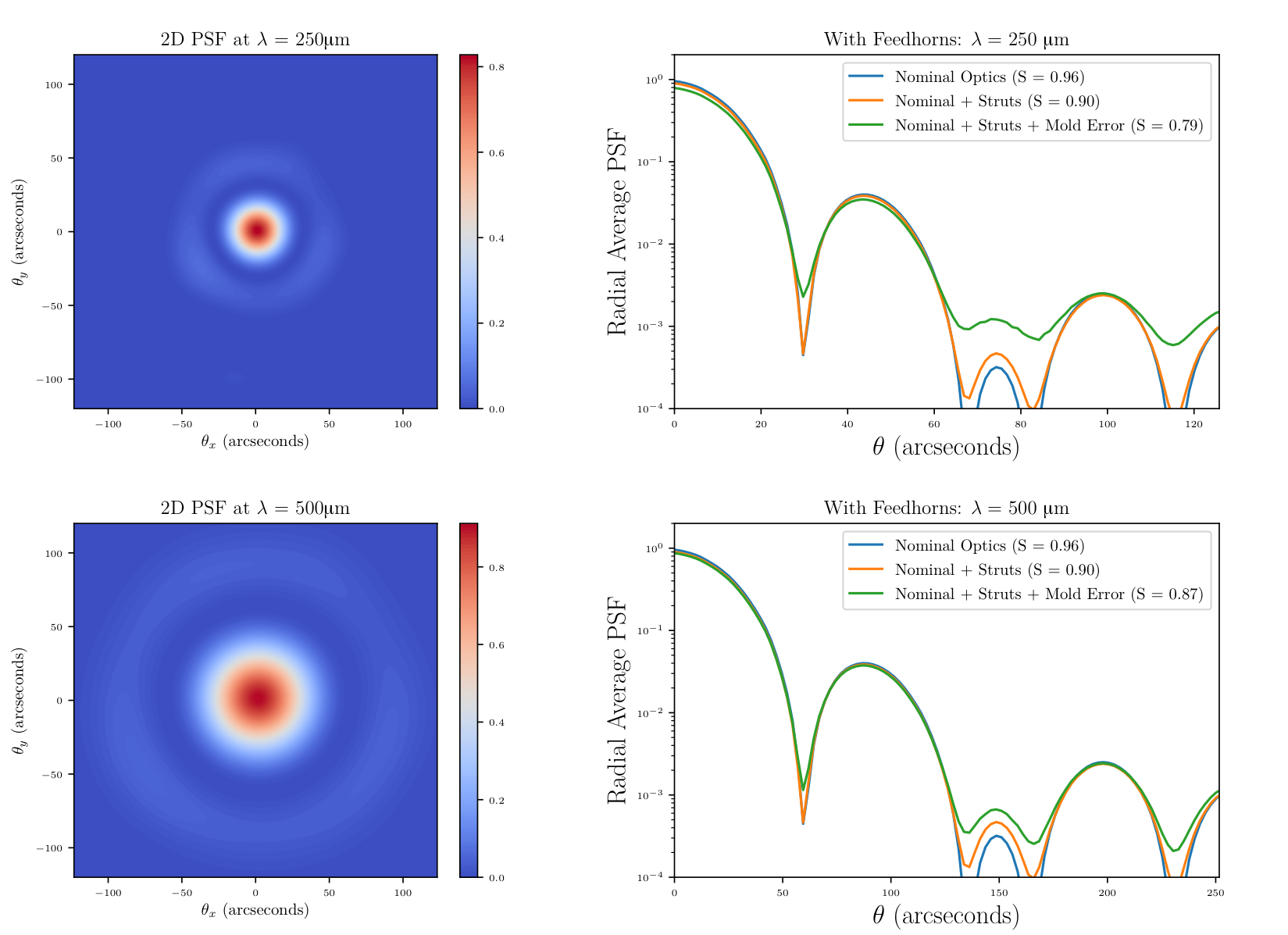}
\caption[Modeled System Point-Spread Function] {Modeled point-spread function for the BLAST-TNG optical system at 250 and 500 \um, based on reference optical prescription (``Nominal Optics"), the telescope obscuration (``Nominal + Struts"), and the measured surface error of the primary mirror mold (``Nominal + Struts + Mold"). The model includes the measured response of the Gaussian illumination of the Lyot stop by the feedhorns.}
\label{figure:psf}
\end{figure}%

The predicted point-spread functions (PSF) of the system at 250 and 500 \um are plotted in Fig. \ref{figure:psf}. Because the Lyot stop underilluminates the primary mirror, the resolution is set by the image of the stop on the primary mirror. The predicted Strehl ratio of the system exceeds the 80\% convention for the diffraction limit based on the Mar\'echal condition across all but the shortest wavelengths. The anticipated WFE of the system is low enough that despite a reduction of power in the main beam, the resolution is not degraded beyond the diffraction limit. A summary of the expected optical performance is given in Table \ref{table:psf}.

% Table generated by Excel2LaTeX from sheet 'Expected Strehl and Resolution'
\begin{table}[htbp]
  \centering
  \caption{Expected Telescope Performance Summary}
    \begin{tabular}{ccc}
    \toprule
    \textbf{Wavelength} & \textbf{Beam FWHM} & \textbf{Strehl Ratio} \\
    \midrule
    \midrule
    250 \um & 30\arcsec & 0.79 \\
    350 \um & 41\arcsec & 0.84 \\
    500 \um & 59\arcsec & 0.87 \\
    \bottomrule
    \end{tabular}%
  \label{table:psf}%
\end{table}%

\section{Balloon-Borne Telescope Platform}
\label{section:gondola}
The telescope is mounted in an altazimuth configuration on a pointed gondola which is suspended from the balloon flight train. The gondola comprises two major mechanical systems: an outer frame which allows the telescope to rotate in azimuth, and an inner frame that can be precisely pointed in elevation with respect to the outer frame by way of a direct-drive motor. The outer frame is suspended by four steel cables from the a high-torque pivot motor which attaches to the base of the balloon flight train. Primary azimuth pointing is achieved by a high moment of inertia reaction wheel which provides fine control of the scan speed, but saturates if large slews or fast azimuth scans are required. The pivot motor servos off the reaction wheel speed to provide coarse pointing during slews and dump angular momentum from the reaction wheel to the balloon flight train.

BLAST-TNG operates primarily in a continuous raster-scan mode, with faster ($\sim$0.5\deg/s) scans in azimuth and slow (\textless 0.1$^\circ$/s) drifts in elevation. Data taken during azimuth turnarounds is discarded due to these vibrationally-induced thermal instabilities. Typical observations will map square or circular regions on the sky of sizes from a $\sim$5-50 square degrees, with the lower limit set by inefficiencies in the observation-to-turnaround time, and the upper limit set by the scan speed and the 1/f knee of the detectors. 

A suite of pointing sensors are used to continuously measure the attitude, geographic location, and trajectory of the payload and are read in by the flight computers to calculate the real-time telescope pointing solution in right ascension (RA) and declination (DEC). Absolute pointing information of the telescope is primarily provided by two autonomous daytime-operating star cameras, mounted on a carbon fiber truss above the cryostat, pointed parallel to the optical axis. These cameras contain a high-resolution integrating CCD camera controlled by a single-board computer, both mounted in an aluminum pressure vessel. The camera observes a 2\deg by 2.5\deg area, and the exposure time, aperture, and focus can be controlled by the single-board computer and stepper motors mounted to the camera lens \cite{rex_starcams}. The control computer runs the attitude-determination program \texttt{STARS} developed for the EBEX experiment \cite{joy_attitude, chapman_stars}. However, the typical scan speed for BLAST-TNG is around ~0.5 $^\circ$/s, meaning that to avoid blurring of the images, the star cameras can only capture images at the turnarounds of the scans, which are a few seconds apart. Two three-axis fiber-optic gyroscope units \footnote{KVH Industries DSP-1760} mounted on opposite sides of the inner frame are used to precisely measure the angular velocity of the inner frame and interpolate the telescope pointing in between star camera solutions. These sensors are able to compute the solution to \textless 5\arcmin \space rms during flight, and to \textless 5\arcsec \space rms after post-flight pointing reconstruction \cite{gandilo_attitude}.
\begin{figure}[htb!]
\begin{center}
\includegraphics[height=.5\textheight]{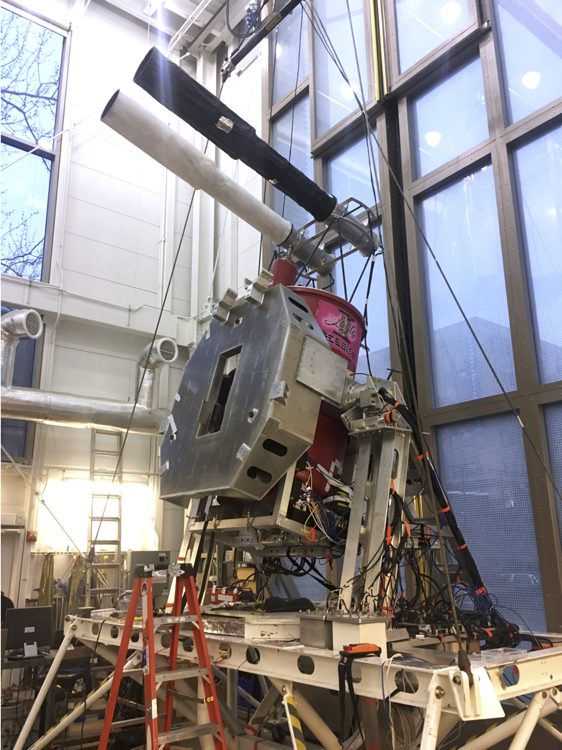} 
\caption[Cryostat on Inner Frame]{Photograph of the BLAST-TNG gondola, showing the telescope mounting surface on the front of the inner frame, the cryogenic receiver (painted red), and the two autonomous daytime star cameras mounted above the receiver. Each star camera has a long baffle to reject stray light and reflections off of the telescope baffle. For flight, both baffles will be painted white to reduce their thermal emissivity.}
\label{figure:cryo_on_gondola}
\end{center}
\end{figure}

\subsection{Mechanical Requirements}\label{subsection:MechReq}
The balloon-borne platform presents a distinct set of constraints and requirements, combining the gravitational stresses of a ground-based telescope, and the mass constraints and extreme thermal environment of a space telescope. For a given size balloon, the altitude of the balloon during flight is determined by the buoyancy of the helium in the balloon and the mass of the payload. To reach the required altitudes with the 34 million cubic feet balloon used by BLAST-TNG the this places a absolute maximum mass requirement of 3,600 kg. Reducing the mass as much as possible below this maximum reduces both the structural demands on the gondola and suspension elements and the torque output from the pointing motors. For BLAST-TNG an upper limit of 3,200 kg was set on the total payload mass, though we expect the actual mass of the payload to be closer to 2,700 kg. 

The main mechanical requirement of the inner frame is that it maintain precise alignment between the telescope, the receiver, and the star cameras. Any relative motion of the telescope with respect to the receiver will cause spurious blurring and streaking of the images. Equally important is that the telescope beam not move with respect to the star cameras, as this would cause an elevation-dependent systematic pointing offset in the maps and jitter during elevation turnarounds.  The absolute offset angle between the star camera beam and the telescope beam is not critical, as long as they are roughly aligned such that the star cameras have a clear view of the sky at all elevation angles. This means the star cameras must be aligned to the telescope bore sight to within a few degrees. Misalignment between the telescope and the receiver shifts the center of the beam on the focal planes, leading to under-illumination of the pixels at the edge of the array. To achieve the necessary rigidity of the inner frame, we required that the relative pointing misalignment between the telescope and the receiver due to deformation of the mounting structure be less than half the FWHM of the beam at the smallest wavelength of observation over the full range of observed elevations between 20 and 60 degrees.

%%%%%%%%%%%%%%%%%%%%%%%%%%%%%%%%%%%%%%%%%%%%%%%%%%%%%%%%%%%%%%%%%%%%%%%%%%%%%%
\subsection{Telescope Mount Design}
%%%%%%%%%%%%%%%%%%%%%%%%%%%%%%%%%%%%%%%%%%%%%%%%%%%%%%%%%%%%%%%%%%%%%%%%%%%%%%

To keep the BLAST-TNG inner frame stiff and compact, while separating the bending stresses from the telescope optics, the frame is composed of two joined sections: a ring of alumnum c-channel which attaches around the waist of the cryostat, and a  welded monocoque structure of thin aluminum sheets which forms a broad flat surface for the telescope to mount at the front of the frame. A cut-out in this front plate allows the telescope beam to pass through the inner frame, and affords access to the cryostat front window.  The cryostat readout electronics are mounted securely to a welded rack hanging below the rear section of the inner frame. Aluminum 6061-T6 was selected for the frame material because it is easily machined, welded, relatively inexpensive, and there are many readily available standard structural beams and other elements that can be incorporated into the design. Because aluminum is also a good thermal conductor, the frame can be used as a heat sink for the flight electronics.

The monocoque structure contains four internal ribs, thin vertical sheets of metal, in addition to thick sidewalls which make it very stiff along the direction of the optical axis to minimize sag from the weight of the telescope. The location of the internal webbing structure is shown in Fig. \ref{figure:sag}. To reduce mass, most panels include oval cut-outs where their presence would not unacceptably reduce the structural strength or interfere with the mounting locations of the mirror, motors, or electronics. A series of 13 mm-thick pads were welded to the front facesheet for mounting the telescope. Before joining the front and rear frame sections, these pads were machined flat and coplanar after they were welded on in a machine large enough to accommodate the full monocoque assembly. Only after surfacing all these mounts were the bolt holes tapped for the telescope mount. This process ensured that the telescope would mount on a flat surface which would not stress the optics.

%%%%%%%%%%%%%%%%%%%%%%%%%%%%%%%%%%%%%%%%%%%%%%%%%%%%%%%%%%%%%%%%%%%%%%%%%%%%%%
\subsection{Modeling Pointing Performance}
%%%%%%%%%%%%%%%%%%%%%%%%%%%%%%%%%%%%%%%%%%%%%%%%%%%%%%%%%%%%%%%%%%%%%%%%%%%%%%

In order to size the structural members of the inner frame, a detailed finite element model (FEM) was developed and a series of trade studies performed to achieve the desired stiffness of the frame with as low a mass as possible. A number of parameters were varied in these studies, including the monocoque web thickness, the height and web thickness of the rear c-channel, the thickness of the top (cryostat mount) and bottom rear plates, the cryostat support gussets, geometry of the front and rear sections, and mesh size. 

To quickly evaluate the results of the FEM simulations of the inner frame assembly without running a full opto-mechanical analysis of the telescope and mount, the pointing offset between the receiver and the telescope optics were estimated for each simulation. The x, y, and z coordinates of of five finite elements were tracked before and after running the simulations. Two elements at the front and back of the cryostat top plate were used to define the cryostat boresight vector, $\vec{V}_{cryo}$. Three elements on the telescope optics bench, at the rear of the three secondary mirror struts, were used to define a plane perpendicular to the optical axis of the telescope. The normal vector to this optics bench plane, $\vec{N}_{OB}$ points along the telescope optical axis. The pointing offset, $theta$, was defined as angle between these two vectors: $\cos(\theta) = \vec{N}_{OB} \bullet \vec{V}_{cryo} $.

Because this bending-induced pointing offset changes with elevation, the FEM simulation was run at 20\deg \space and 60\deg, the upper and lower elevation limits during flight, and  the differential pointing offset, $\Delta \theta$, between these angles was calculated. 

\begin{figure}[ht!]
\begin{center}
\includegraphics[width=\textwidth]{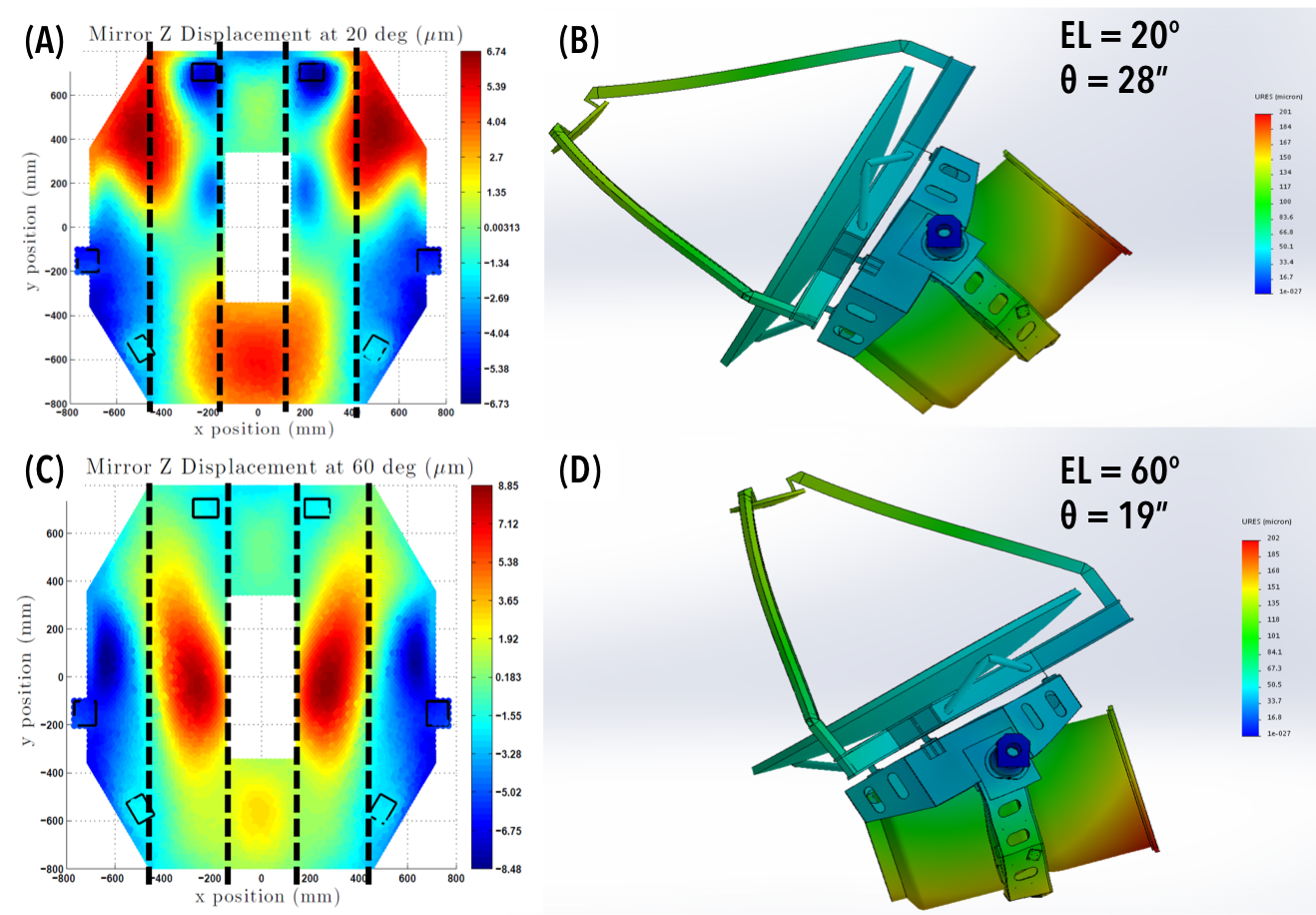} 
\caption[Inner Frame FEM Results]{Results of finite element modeling of bending of the inner frame under 1-g gravitaional loading at different elevation angles. The deformation scale is exaggerated by several hundred times in order to make the direction of the bending more clear. The largest displacement in the model is the cryostat, and is only 200 \um. Figures A and C show the calculated bending in z-direction of the front facesheet of the inner frame monocoque. The six rectangles around the perimeter are the telescope mount points. The four vertical dashed lines show the location of the internal ribs of the monocoque. As expected most of the bending occurs between the ribs. Figures B and C show the total displacement from bending of the inner frame, cryostat, and simplified telescope.}
\label{figure:sag}
\end{center}
\end{figure}

Because the simulations were meant to evaluate the stiffness of the inner frame itself, not all elements in the assembly were modeled. Only the exterior of the cryostat was included, and the internal components were not modeled. The internal optical components have been modeled separately \cite{tyr_thesis}. The star cameras and their mount were not included in these simulations, as we required the star camera mount be stiff enough that any bending would be subdominant to the bending of the inner frame. The aluminum flexures that attach the telescope optics bench to the front of the inner frame were modeled as perfectly rigid elements as the details of their design were still being developed at the manufacturer. This assumption means that the simulations likely over-estimate the bending of the front facesheet of the inner frame. The Fig. \ref{figure:sag} shows the deformation of the front facesheet of the monocoque from the simulations at the upper and lower limits of the elevation range. The magnitude of the deformation across the full face is less than 10 \um. The mean offset in z was calculated for each of the six telescope mount points at each elevation, and these offsets were put into the detailed telescope FEM model developed at Alliance Spacesystems. The displacements of the mount points had no observed effect on the primary mirror bending or WFE. 

The results of the FEM simulations of the as-built inner frame are shown in Fig. \ref{figure:sag}. At an elevation of 30\deg, the pointing offset is 28\arcsec, and at an elevation of 60\deg, the pointing offset is 19\arcsec, giving a differential pointing offset over range of observing elevations of 9\arcsec. This is less than half the beam FWHM at the shortest observed wavelength. This offset is of roughly the same order as the anticipated differential pointing offset due to bending of the telescope optics alone. Even if the bending of the telescope optics causes the beam to move in the same direction as the bending of the inner frame, we expect that the elevation-dependent pointing offset from bending/motion of the telescope optics, inner frame, and cryostat to be less than the size of the beam at 250 \si{\um}.

\subsection{Thermal Environment}
\label{section:thermal}

The BLAST-TNG payload thermal environment is controlled by extensive baffling and Sun shields. The thermal environment of the telescope and the payload must be carefully controlled to avoid direct solar illumination of the optical system, and to ensure all components operate within allowable temperature ranges. The flight trajectory and the conditions during launch, ascent, and descent are largely unpredictable, and can vary widely between different launches. As such, it is necessary to design the gondola platform to handle a wide range conditions and stresses. 

The Sun shield design follows the approach from the BLASTPol experiment, detailed in Soler et. al., 2014 \cite{soler_thermal}. A 6.5 m long aluminized mylar baffle surrounds the telescope optics. The baffle is formed around a truss of carbon fiber tubes \footnote{CST Composites} bonded to aluminum fittings, incorporating design elements from the X-Calibur gondola \cite{xcalibur}. An outer Sun shield built from welded aluminum pipe\footnote{GSM Industrial Inc.} encloses the entire payload and acts as a ground screen protecting the entire instrument from both direct and reflected solar illumination. A detailed ThermalDesktop \footnote{Cullimore and Ring Inc.} model was used to determine the placement of the reflective panels and predict the operating temperatures of the major systems and electronic components.

\begin{figure}[ht!]
\begin{center}
\includegraphics[width=.8\textwidth]{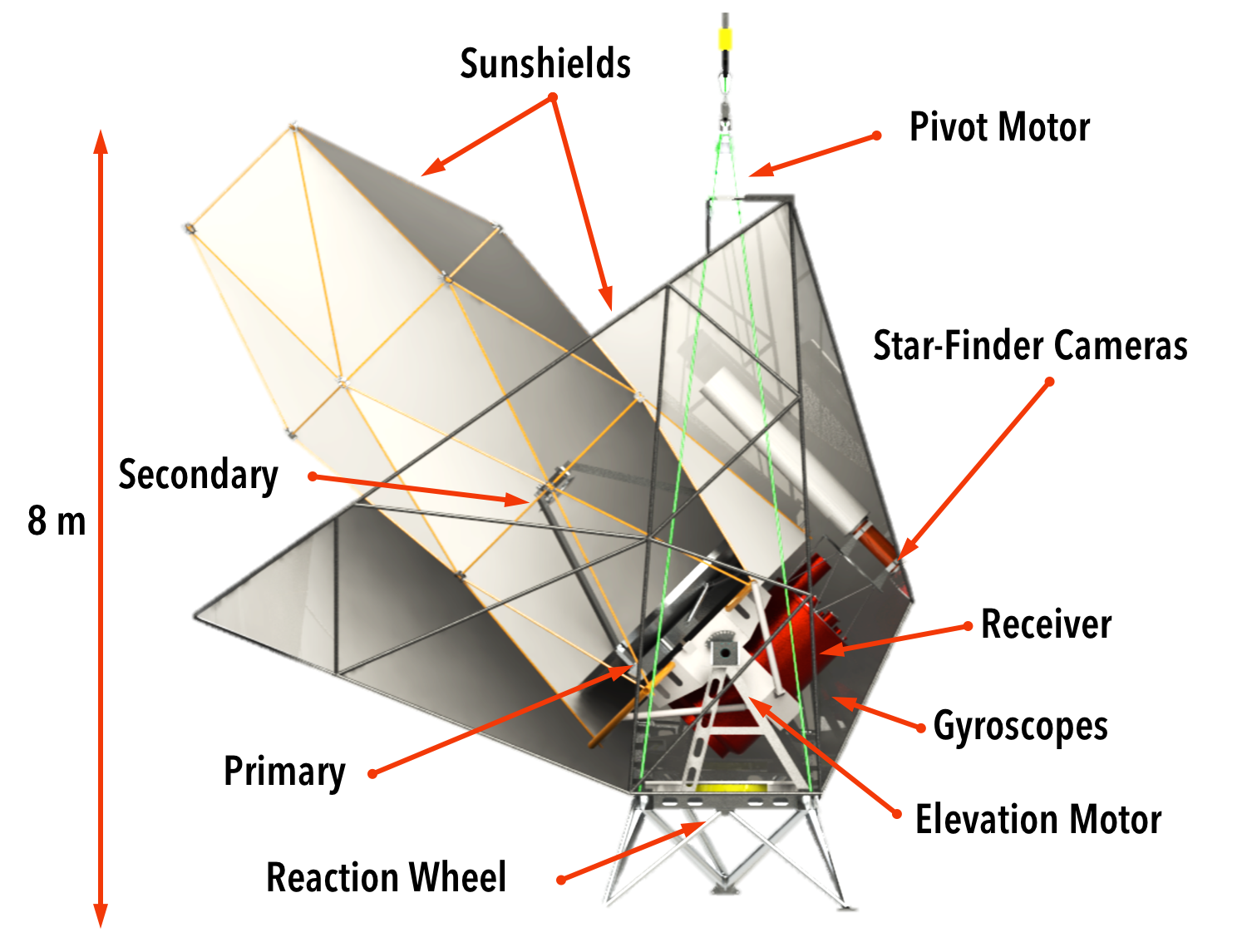} 
\caption[Render of BLAST-TNG with Critical Components Labeled ]{Render of the BLAST-TNG payload, with critical components labeled, including the telescope and pointed gondola platform. The cryogenic receiver, shown in red, supports the two boresight star cameras. The large asymmetric Sun shields enclose the payload, protecting the components from solar illumination and Earth-shine, and allow the gondola to point to within 35\deg in azimuth from the Sun on the starboard side.}
\label{figure:gondola_redline}
\end{center}
\end{figure}

The visibility of different regions of the sky throughout the flight is set by the geometry of the telescope baffle. The telescope pointing scheduler is constrained by limits determined from the thermal model based on the geometry of the baffle and the position of the Sun. At every elevation angle, the allowed azimuth angles are determined by ensuring that there is no direct thermal illumination of the telescope primary or secondary mirrors. Because the balloon may in latitude from its launch location depending on stratospheric wind currents, these constraints are calculated at extreme cases of \textpm10\deg latitude drifts, and recalculated at different dates throughout the anticipated flight. To allow for pointing towards certain high-priority targets, the baffle is designed to allow pointing to within approximately 35\deg of the sun at most elevation angles throughout the flight.

\section{Conclusion}

The BLAST-TNG experiment features one of the most technologically ambitious telescopes ever flown on a balloon experiment. With a 2.5 m diameter carbon fiber composite primary mirror, BLAST-TNG will be able to map the submillimeter polarized thermal emission from interstellar dust at sub-arcminute resolution, probing the magnetic field structure of molecular clouds and the diffuse interstellar medium on previously unexplored angular scales. The telescope has been completed and will be integrated with the balloon gondola and cryogenic receiver during pre-flight systems integration at the NASA Columbia Scientific Ballooning Facility in Palestine, TX, in preparation for a planned 28 day stratospheric balloon flight from McMurdo Station, Antarctica during the winter of 2018/2019. 

\acknowledgments % equivalent to \section*{ACKNOWLEDGMENTS}       
 
The BLAST-TNG collaboration acknowledges the support of NASA under award numbers NNX13AE50G and 80NSSC18K0481, and the NNX13CM03C. Detector development is supported in part by NASA through NNH13ZDA001N-APRA. J.D.S. acknowledges the support from the European Research Council (ERC) under the Horizon 2020 Framework Program via the Consolidator Grant CSF-648505. S.G. is supported through a NASA Earth and Space Science Fellowship (NESSF)  NNX16AO91H. The BLAST-TNG telescope is supported in part through the NASA SBIR/STTR office and developed at Alliance Spacesystems. The collaboration also acknowledges the extensive machining, design, and fabrication efforts of Jeffrey Hancock and Harold Borders at the University of Pennsylvania and Matthew Underhill at Arizona State University. Paul Dowkontt at Washington University in St. Louis provided design assistance for the sun shield fitting design. The BLAST-TNG team also recognizes the contribution of undergraduate and post-baccalaureate interns to the gondola development, especially Mark	Giovinazzi, Erin Healy, Gregory Kofman, Aaron Mathews, Timothy McSorely, Michael Plumb, Steven Russel, and Nathan Schor.

% References
\bibliography{report} % bibliography data in report.bib

\begin{thebibliography}{10}

\bibitem{enzo_blast}
{Pascale}, E., {Ade}, P.~A.~R., {Bock}, J.~J., {Chapin}, E.~L., {Chung}, J.,
  {Devlin}, M.~J., {Dicker}, S., {Griffin}, M., {Gundersen}, J.~O., {Halpern},
  M., {Hargrave}, P.~C., {Hughes}, D.~H., {Klein}, J., {MacTavish}, C.~J.,
  {Marsden}, G., {Martin}, P.~G., {Martin}, T.~G., {Mauskopf}, P.,
  {Netterfield}, C.~B., {Olmi}, L., {Patanchon}, G., {Rex}, M., {Scott}, D.,
  {Semisch}, C., {Thomas}, N., {Truch}, M.~D.~P., {Tucker}, C., {Tucker},
  G.~S., {Viero}, M.~P., and {Wiebe}, D.~V., ``{The Balloon-borne Large
  Aperture Submillimeter Telescope: BLAST},'' {\em \apj}~{\bf 681},  400--414
  (July 2008).

\bibitem{fissel_blastpol}
{Fissel}, L.~M., {Ade}, P.~A.~R., {Angil{\`e}}, F.~E., {Ashton}, P., {Benton},
  S.~J., {Devlin}, M.~J., {Dober}, B., {Fukui}, Y., {Galitzki}, N., {Gandilo},
  N.~N., {Klein}, J., {Korotkov}, A.~L., {Li}, Z.-Y., {Martin}, P.~G.,
  {Matthews}, T.~G., {Moncelsi}, L., {Nakamura}, F., {Netterfield}, C.~B.,
  {Novak}, G., {Pascale}, E., {Poidevin}, F., {Santos}, F.~P., {Savini}, G.,
  {Scott}, D., {Shariff}, J.~A., {Diego Soler}, J., {Thomas}, N.~E., {Tucker},
  C.~E., {Tucker}, G.~S., and {Ward-Thompson}, D., ``{Balloon-Borne
  Submillimeter Polarimetry of the Vela C Molecular Cloud: Systematic
  Dependence of Polarization Fraction on Column Density and Local
  Polarization-Angle Dispersion},'' {\em \apj}~{\bf 824},  134 (June 2016).

\bibitem{planck_XXXII}
{Planck Collaboration}, ``{Planck intermediate results. XXXII. The relative
  orientation between the magnetic field and structures traced by interstellar
  dust},'' {\em \aap}~{\bf 586},  A135 (Feb. 2016).

\bibitem{planck_XXX}
{Planck Collaboration}, ``{Planck intermediate results. XXX. The angular power
  spectrum of polarized dust emission at intermediate and high Galactic
  latitudes},'' {\em \aap}~{\bf 586},  A133 (Feb. 2016).

\bibitem{soler_blastpol}
{Soler}, J.~D., {Ade}, P.~A.~R., {Angil{\`e}}, F.~E., {Ashton}, P., {Benton},
  S.~J., {Devlin}, M.~J., {Dober}, B., {Fissel}, L.~M., {Fukui}, Y.,
  {Galitzki}, N., {Gandilo}, N.~N., {Hennebelle}, P., {Klein}, J., {Li}, Z.-Y.,
  {Korotkov}, A.~L., {Martin}, P.~G., {Matthews}, T.~G., {Moncelsi}, L.,
  {Netterfield}, C.~B., {Novak}, G., {Pascale}, E., {Poidevin}, F., {Santos},
  F.~P., {Savini}, G., {Scott}, D., {Shariff}, J.~A., {Thomas}, N.~E.,
  {Tucker}, C.~E., {Tucker}, G.~S., and {Ward-Thompson}, D., ``{The relation
  between the column density structures and the magnetic field orientation in
  the Vela C molecular complex},'' {\em \aap}~{\bf 603},  A64 (July 2017).

\bibitem{hill}
{Hill}, T., {Motte}, F., {Didelon}, P., {Bontemps}, S., {Minier}, V.,
  {Hennemann}, M., {Schneider}, N., {Andr{\'e}}, P., {Men'shchikov}, A.,
  {Anderson}, L.~D., {Arzoumanian}, D., {Bernard}, J.-P., {di Francesco}, J.,
  {Elia}, D., {Giannini}, T., {Griffin}, M.~J., {K{\"o}nyves}, V., {Kirk}, J.,
  {Marston}, A.~P., {Martin}, P.~G., {Molinari}, S., {Nguyen Luong}, Q.,
  {Peretto}, N., {Pezzuto}, S., {Roussel}, H., {Sauvage}, M., {Sousbie}, T.,
  {Testi}, L., {Ward-Thompson}, D., {White}, G.~J., {Wilson}, C.~D., and
  {Zavagno}, A., ``{Filaments and ridges in Vela C revealed by Herschel: from
  low-mass to high-mass star-forming sites},'' {\em \aap}~{\bf 533},  A94
  (Sept. 2011).

\bibitem{soler_sims_2013}
{Soler}, J.~D., {Hennebelle}, P., {Martin}, P.~G., {Miville-Desch{\^e}nes},
  M.-A., {Netterfield}, C.~B., and {Fissel}, L.~M., ``{An Imprint of Molecular
  Cloud Magnetization in the Morphology of the Dust Polarized Emission},'' {\em
  \apj}~{\bf 774},  128 (Sept. 2013).

\bibitem{guillet_2018}
{Guillet}, V., {Fanciullo}, L., {Verstraete}, L., {Boulanger}, F., {Jones},
  A.~P., {Miville-Desch{\^e}nes}, M.-A., {Ysard}, N., {Levrier}, F., and
  {Alves}, M., ``{Dust models compatible with Planck intensity and polarization
  data in translucent lines of sight},'' {\em \aap}~{\bf 610},  A16 (Feb.
  2018).

\bibitem{draine_hensley}
{Draine}, B.~T. and {Hensley}, B., ``{Magnetic Nanoparticles in the
  Interstellar Medium: Emission Spectrum and Polarization},'' {\em \apj}~{\bf
  765},  159 (Mar. 2013).

\bibitem{cmb_s4}
{Abazajian}, K.~N., {Adshead}, P., {Ahmed}, Z., {Allen}, S.~W., {Alonso}, D.,
  {Arnold}, K.~S., {Baccigalupi}, C., {Bartlett}, J.~G., {Battaglia}, N.,
  {Benson}, B.~A., {Bischoff}, C.~A., {Borrill}, J., {Buza}, V., {Calabrese},
  E., {Caldwell}, R., {Carlstrom}, J.~E., {Chang}, C.~L., {Crawford}, T.~M.,
  {Cyr-Racine}, F.-Y., {De Bernardis}, F., {de Haan}, T., {di Serego
  Alighieri}, S., {Dunkley}, J., {Dvorkin}, C., {Errard}, J., {Fabbian}, G.,
  {Feeney}, S., {Ferraro}, S., {Filippini}, J.~P., {Flauger}, R., {Fuller},
  G.~M., {Gluscevic}, V., {Green}, D., {Grin}, D., {Grohs}, E., {Henning},
  J.~W., {Hill}, J.~C., {Hlozek}, R., {Holder}, G., {Holzapfel}, W., {Hu}, W.,
  {Huffenberger}, K.~M., {Keskitalo}, R., {Knox}, L., {Kosowsky}, A., {Kovac},
  J., {Kovetz}, E.~D., {Kuo}, C.-L., {Kusaka}, A., {Le Jeune}, M., {Lee},
  A.~T., {Lilley}, M., {Loverde}, M., {Madhavacheril}, M.~S., {Mantz}, A.,
  {Marsh}, D.~J.~E., {McMahon}, J., {Meerburg}, P.~D., {Meyers}, J., {Miller},
  A.~D., {Munoz}, J.~B., {Nguyen}, H.~N., {Niemack}, M.~D., {Peloso}, M.,
  {Peloton}, J., {Pogosian}, L., {Pryke}, C., {Raveri}, M., {Reichardt}, C.~L.,
  {Rocha}, G., {Rotti}, A., {Schaan}, E., {Schmittfull}, M.~M., {Scott}, D.,
  {Sehgal}, N., {Shandera}, S., {Sherwin}, B.~D., {Smith}, T.~L., {Sorbo}, L.,
  {Starkman}, G.~D., {Story}, K.~T., {van Engelen}, A., {Vieira}, J.~D.,
  {Watson}, S., {Whitehorn}, N., and {Kimmy Wu}, W.~L., ``{CMB-S4 Science Book,
  First Edition},'' {\em ArXiv e-prints}  (Oct. 2016).

\bibitem{caldwell_EEBB}
Caldwell, R.~R., Hirata, C., and Kamionkowski, M., ``Dust-polarization maps and
  interstellar turbulence,'' {\em The Astrophysical Journal}~{\bf 839}(2),  91
  (2017).

\bibitem{ghosh}
{Ghosh}, T., {Boulanger}, F., {Martin}, P.~G., {Bracco}, A., {Vansyngel}, F.,
  {Aumont}, J., {Bock}, J.~J., {Dor{\'e}}, O., {Haud}, U., {Kalberla},
  P.~M.~W., and {Serra}, P., ``{Modelling and simulation of large-scale
  polarized dust emission over the southern Galactic cap using the GASS Hi
  data},'' {\em \aap}~{\bf 601},  A71 (May 2017).

\bibitem{ebex}
Oxley, P., ``The {EBEX} experiment,'' {\em Proc. SPIE}~{\bf 4857} (2004).

\bibitem{sofia}
Bittner, H. et~al., ``{SOFIA} primary mirror assembly: Structural properties
  and optical performance,'' {\em Proc. SPIE}~{\bf 4857} (2003).

\bibitem{jwst}
Gutro, R., ``James {W}ebb space telescope's beryllium mirrors,'' {\em
  http://www.nasa.gov/topics/technology/feature/webb-beryllium.html}  (2013).

\bibitem{catanzaro_blast05}
{Catanzaro}, B.~E., {Pham}, T., {Olmi}, L., {Martinson}, K.~E., and {Devlin},
  M., ``"design and fabrication of a lightweight 2-m telescope for the
  balloon-borne large-aperture submillimeter telescope: Blast",'' {\em
  Proc.SPIE}~{\bf 4818},  4818 -- 4818 -- 8 (2002).

\bibitem{potter}
Potter, P., ``A new horn antenna with suppressed sidelobes and equal
  bandwidths,'' (1963).

\bibitem{lourie_cryo}
{Lourie}, N.~P., {Ade}, P., {Angil{\`e}}, F.~E., {Ashton}, P., {Austermann},
  J., {Devlin}, M., {Dober}, B.~J., {Fissel}, L.~M., {Gao}, J., {Gordon}, S.,
  {Groppi}, C.~E., {Hilton}, G.~C., {Hubmayr}, J., {Klein}, J., {Li}, D.,
  {Lowe}, I., {Mani}, H., {Mauskopf}, P., {McKenney}, C., {Nati}, F., {Novak},
  G., {Pascale}, E., {Pisano}, G., {Sinclair}, A., {Soler}, J.~D., {Tucker},
  C., {Vissers}, M., and {Williams}, P., ``{Preflight characterization of the
  BLAST-TNG receiver and detector arrays},'' in [{\em Millimeter,
  Submillimeter, and Far-Infrared Detectors and Instrumentation for Astronomy
  IX}{\nolinebreak\hspace{0.1em}]},  {\em \procspie} {\bf 10708} (2018).

\bibitem{tyr_blasttng_spie}
{Galitzki}, N., {Ade}, P., {Angil{\`e}}, F.~E., {Ashton}, P., {Austermann}, J.,
  {Billings}, T., {Che}, G., {Cho}, H.-M., {Davis}, K., {Devlin}, M., {Dicker},
  S., {Dober}, B.~J., {Fissel}, L.~M., {Fukui}, Y., {Gao}, J., {Gordon}, S.,
  {Groppi}, C.~E., {Hillbrand}, S., {Hilton}, G.~C., {Hubmayr}, J., {Irwin},
  K.~D., {Klein}, J., {Li}, D., {Li}, Z.-Y., {Lourie}, N.~P., {Lowe}, I.,
  {Mani}, H., {Martin}, P.~G., {Mauskopf}, P., {McKenney}, C., {Nati}, F.,
  {Novak}, G., {Pascale}, E., {Pisano}, G., {Santos}, F.~P., {Scott}, D.,
  {Sinclair}, A., {Soler}, J.~D., {Tucker}, C., {Underhill}, M., {Vissers}, M.,
  and {Williams}, P., ``{Instrumental performance and results from testing of
  the BLAST-TNG receiver, submillimeter optics, and MKID detector arrays},'' in
  [{\em Millimeter, Submillimeter, and Far-Infrared Detectors and
  Instrumentation for Astronomy VIII}{\nolinebreak\hspace{0.1em}]},  {\em
  \procspie} {\bf 9914},  99140J (July 2016).

\bibitem{moisture}
{Augl}, J.~M. and {Berger}, A.~E., ``The effect of moisture on carbon fiber
  reinforced epoxy composites i. diffusion,'' (1976).

\bibitem{catanzaro_blast_metrology}
{Catanzaro}, B.~E., {Thomas}, J.~A., {Small}, D.~W., {Johnston}, R.~A.,
  {Barber}and Steven J.~{Connell}, D.~D., and Shaun A.~{Whitmore}, E. J.~C.,
  ``Optical metrology for testing an all-composite 2-m-diameter mirror,'' {\em
  Proc.SPIE}~{\bf 4444},  4444 -- 4444 -- 17 (2001).

\bibitem{wmap_optics}
Page, L., Jackson, C., Barnes, C., Bennett, C., Halpern, M., Hinshaw, G.,
  Jarosik, N., Kogut, A., Limon, M., Meyer, S.~S., Spergel, D.~N., Tucker,
  G.~S., Wilkinson, D.~T., Wollack, E., and Wright, E.~L., ``The optical design
  and characterization of the microwave anisotropy probe,'' {\em The
  Astrophysical Journal}~{\bf 585}(1),  566 (2003).

\bibitem{maven}
Connoly, D., ``Maven high-gain antenna,'' (April 2012).
\newblock [Online; posted 23-April-2012].

\bibitem{shack_hartmann_jwst}
{Kiikka}, C., {Neal}, D.~R., {Kincade}, J., {Bernier}, R., {Hull}, T.,
  {Chaney}, D., {Farrer}, S., {Dixson}, J., {Causey}, A., and {Strohl}, S.,
  ``The jwst infrared scanning shack hartman system: a new in-process way to
  measure large mirrors during optical fabrication at tinsley,'' {\em
  Proc.SPIE}~{\bf 6265},  6265 -- 6265 -- 11 (2006).

\bibitem{laser_tracker}
{Burge}, J.~H., {Su}, P., {Zhao}, C., and {Zobrist}, T., ``Use of a commercial
  laser tracker for optical alignment,'' {\em Proc.SPIE}~{\bf 6676},  6676 --
  6676 -- 12 (2007).

\bibitem{rex_starcams}
{Rex}, M., {Chapin}, E., {Devlin}, M.~J., {Gundersen}, J., {Klein}, J.,
  {Pascale}, E., and {Wiebe}, D., ``{BLAST autonomous daytime star cameras},''
  in [{\em Society of Photo-Optical Instrumentation Engineers (SPIE) Conference
  Series}{\nolinebreak\hspace{0.1em}]},  {\em \procspie} {\bf 6269},  62693H
  (June 2006).

\bibitem{joy_attitude}
Didier, J., Chapman, D., Aboobaker, A.~M., Araujo, D., Grainger, W., Hanany,
  S., Helson, K., Hillbrand, S., Korotkov, A., Limon, M., Miller, A.,
  Reichborn-Kjennerud, B., Sagiv, I., Tucker, G., and Vinokurov, Y., ``A
  high-resolution pointing system for fast scanning platforms: The ebex
  example,'' in [{\em 2015 IEEE Aerospace
  Conference}{\nolinebreak\hspace{0.1em}]},   1--15 (March 2015).

\bibitem{chapman_stars}
{Chapman}, D., {Didier}, J., {Hanany}, S., {Hillbrand}, S., {Limon}, M.,
  {Miller}, A., {Reichborn-Kjennerud}, B., {Tucker}, G., and {Vinokurov}, Y.,
  ``{STARS: a software application for the EBEX autonomous daytime star
  cameras},'' in [{\em Software and Cyberinfrastructure for Astronomy
  III}{\nolinebreak\hspace{0.1em}]},  {\em \procspie} {\bf 9152},  915212 (July
  2014).

\bibitem{gandilo_attitude}
{Gandilo}, N.~N., {Ade}, P.~A.~R., {Amiri}, M., {Angil{\`e}}, F.~E., {Benton},
  S.~J., {Bock}, J.~J., {Bond}, J.~R., {Bryan}, S.~A., {Chiang}, H.~C.,
  {Contaldi}, C.~R., {Crill}, B.~P., {Devlin}, M.~J., {Dober}, B., {Dor{\'e}},
  O.~P., {Farhang}, M., {Filippini}, J.~P., {Fissel}, L.~M., {Fraisse}, A.~A.,
  {Fukui}, Y., {Galitzki}, N., {Gambrel}, A.~E., {Golwala}, S., {Gudmundsson},
  J.~E., {Halpern}, M., {Hasselfield}, M., {Hilton}, G.~C., {Holmes}, W.~A.,
  {Hristov}, V.~V., {Irwin}, K.~D., {Jones}, W.~C., {Kermish}, Z.~D., {Klein},
  J., {Korotkov}, A.~L., {Kuo}, C.~L., {MacTavish}, C.~J., {Mason}, P.~V.,
  {Matthews}, T.~G., {Megerian}, K.~G., {Moncelsi}, L., {Morford}, T.~A.,
  {Mroczkowski}, T.~K., {Nagy}, J.~M., {Netterfield}, C.~B., {Novak}, G.,
  {Nutter}, D., {O'Brient}, R., {Pascale}, E., {Poidevin}, F., {Rahlin}, A.~S.,
  {Reintsema}, C.~D., {Ruhl}, J.~E., {Runyan}, M.~C., {Savini}, G., {Scott},
  D., {Shariff}, J.~A., {Soler}, J.~D., {Thomas}, N.~E., {Trangsrud}, A.,
  {Truch}, M.~D., {Tucker}, C.~E., {Tucker}, G.~S., {Tucker}, R.~S., {Turner},
  A.~D., {Ward-Thompson}, D., {Weber}, A.~C., {Wiebe}, D.~V., and {Young},
  E.~Y., ``{Attitude determination for balloon-borne experiments},'' in [{\em
  Ground-based and Airborne Telescopes V}{\nolinebreak\hspace{0.1em}]},  {\em
  \procspie} {\bf 9145},  91452U (July 2014).

\bibitem{tyr_thesis}
Galitzki, N., {\em MAGNETIC FIELDS IN MOLECULAR CLOUDS: THE BLASTPOL AND
  BLAST-TNG EXPERIMENTS}, PhD thesis, University of Pennsylvania (2016).

\bibitem{soler_thermal}
{Soler}, J.~D., {Ade}, P.~A.~R., {Angil{\`e}}, F.~E., {Benton}, S.~J.,
  {Devlin}, M.~J., {Dober}, B., {Fissel}, L.~M., {Fukui}, Y., {Galitzki}, N.,
  {Gandilo}, N.~N., {Klein}, J., {Korotkov}, A.~L., {Matthews}, T.~G.,
  {Moncelsi}, L., {Mroczkowski}, A., {Netterfield}, C.~B., {Novak}, G.,
  {Nutter}, D., {Pascale}, E., {Poidevin}, F., {Savini}, G., {Scott}, D.,
  {Shariff}, J.~A., {Thomas}, N.~E., {Truch}, M.~D., {Tucker}, C.~E., {Tucker},
  G.~S., and {Ward-Thompson}, D., ``{Thermal design and performance of the
  balloon-borne large aperture submillimeter telescope for polarimetry
  BLASTPol},'' in [{\em Ground-based and Airborne Telescopes
  V}{\nolinebreak\hspace{0.1em}]},  {\em \procspie} {\bf 9145},  914534 (July
  2014).

\bibitem{xcalibur}
{Kislat}, F., {Beheshtipour}, B., {Dowkontt}, P., {Guarino}, V., {Lanzi},
  R.~J., {Okajima}, T., {Braun}, D., {Cannon}, S., {de Geronimo}, G.,
  {Heatwole}, S., {Hoorman}, J., {Li}, S., {Mori}, H., {Shreves}, C.~M.,
  {Stuchlik}, D., and {Krawczynski}, H., ``{Design of the Telescope Truss and
  Gondola for the Balloon-Borne X-ray Polarimeter X-Calibur},'' {\em Journal of
  Astronomical Instrumentation}~{\bf 6},  1740003 (2017).

\end{thebibliography}
\bibliographystyle{spiebib} % makes bibtex use spiebib.bst

\end{document}